\documentclass[prl,twocolumn,aps,showpacs,superscriptaddress]{revtex4}
\usepackage{graphics}
\usepackage{epsfig}
\usepackage{amsmath}
\usepackage{amssymb}
\usepackage{bbm}
\usepackage{bm}
\usepackage{color}

\newcommand{\pbar}{\langle p \rangle}

\newcommand{\sbar}{\langle \hat{\sigma} \rangle}
\newcommand{\Abar}{\langle a \rangle}
\newcommand{\Abarp}{\langle a(p) \rangle}

\newcommand{\f}{\boldsymbol{f}}
\newcommand{\bond}{\boldsymbol{r}}
\newcommand{\bigP}{\mathcal{P}}
\newcommand{\A}{\mathcal{A}}

\begin{document}
\title{Entropy maximization in the force network ensemble for granular solids}

\author{Brian P.~Tighe}
\affiliation{Instituut-Lorentz, Universiteit Leiden, Postbus 9506, 2300 RA Leiden, The Netherlands}
\author{Adrianne R.~T.~van Eerd}
\affiliation{Condensed Matter and Interfaces, Debye Institute for NanoMaterials Science, Utrecht University, P.O.~Box 80.000, 3508 TA Utrecht, The Netherlands}
\author{Thijs J.~H.~Vlugt}
\affiliation{Delft University of Technology, Process \& Energy Laboratory, Leeghwaterstraat 44, 2628 CA Delft, The Netherlands}

\date{\today}

\begin{abstract}
A long-standing issue in the area of granular media is the tail of the force distribution, in particular whether this is exponential, Gaussian, or even some other form. Here we resolve the issue for the case of the force network ensemble in two dimensions. We demonstrate that conservation of the total area of a reciprocal tiling, a direct consequence of local force balance, is crucial for predicting the local stress distribution. Maximizing entropy while conserving the tiling area and total pressure leads to a distribution of local pressures with a generically Gaussian tail that is in excellent agreement with numerics, both with and without friction and for two different contact networks.
\end{abstract}

\maketitle

\begin{figure}[tbp] 
\centering
\includegraphics[clip,width=0.75\linewidth]{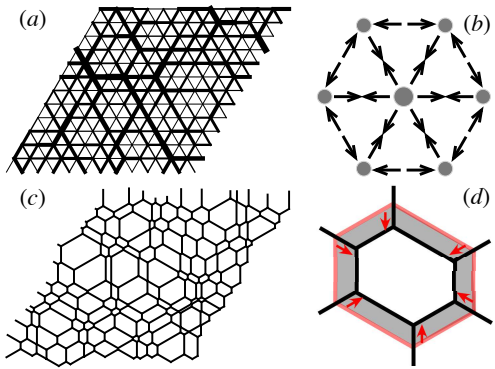}\\
\caption{A force network (a) on the periodic frictionless triangular lattice. Edges represent contact forces; larger forces are thicker. (b) A wheel move. Arrows represent changes to the forces on each grain. (c) The reciprocal tiling corresponding to (a). Larger forces map to longer lines. (d) A move in the tiling. Shaded area is being exchanged among tiles. }
\label{fig:realrecip}
\end{figure}
In a granular system the interactions among individual grains are dissipative \cite{jaeger96}, so that an undriven system eventually jams into a static, mechanically stable configuration \cite{liu98}. There are typically many jammed states consistent with a given set of macroscopic constraints, e.g.~fixed grain number and fixed pressure or volume. A granular packing can move from one jammed configuration to another only under externally imposed fluctuations such as shaking or tapping, because grains are too massive to rearrange thermally \cite{jaeger96}. Despite this nonequilibrium nature, Edwards proposed that the methods of equilibrium statistical mechanics might describe many material properties of jammed media \cite{edwards}. A successful statistical mechanics of jammed systems would represent an important theoretical handle on the physics of granular media. It would, for example, permit the calculation of grain scale statistical properties, e.g.~volumes \cite{volumes} or stresses \cite{stresses,goddard04,henkes07,ngan05}, from a small number of global constraints. 

In this Letter we derive an analytic expression for the probability density $\rho(p)$ of pressures on individual grains in the bulk, which characterizes the strikingly heterogeneous force networks observed in granular solids \cite{trush05}. We work within the force network ensemble of Snoeijer {\em et al}.~\cite{snoeijer04} and confirm our results with highly accurate numerics. The force network ensemble comprises all balanced force configurations with a fixed global stress tensor $\sbar$ on a fixed contact network beyond marginal rigidity. The ensemble is both a minimal model for the statistics of local stresses and a basic test for any statistical mechanical theory of stress states in granular systems. 

There is as yet no clear consensus on the form of the distribution of local stresses in granular media. Of particular interest is the large-stress tail, which early experiments found to be exponential when measured on the boundary \cite{exptboundary}. More recent measurements in the bulk \cite{exptbulk}, along with numerics \cite{numerics}, find distributions that bend downward on a semilogarithmic plot, suggesting faster than exponential decay. A number of proposed theories exploit an analogy to the microcanonical ensemble to arrive at a Boltzmann-like exponential tail \cite{stresses,goddard04,henkes07}. These theories should in principle apply to the force network ensemble, but numerical simulations tailor-made to accurately sample large contact forces unambiguously show a Gaussian tail in the force network ensemble in 2D \cite{vaneerd07}. As the present statistical mechanics approaches fail to describe simple models like the force network ensemble, they must be missing an important ingredient. We argue that {\em local force balance} is absolutely crucial to describe the correct stress statistics. In particular, we show that the pressure distribution in the force network ensemble directly follows from entropy maximization, but {\em only} when it respects a conserved quantity overlooked in previous theories. This conserved quantity follows from force balance at the grain scale, and leads to excellent agreement with numerics for both small and large forces.

% We study the 2D ensemble for both frictionless and frictional packings. We demonstrate that maximization of an entropy subject to two global constraints -- fixed total pressure and fixed area of a reciprocal tiling -- yields the correct distribution of pressures. Earlier ensemble approaches to local stress statistics \cite{stresses,goddard04,henkes07} have included only the pressure constraint; this results in exponential statistics of large stresses. Conservation of the reciprocal tiling area, which results from vector force balance at the grain scale, renders the large-$p$ tail of the pressure distribution generically Gaussian, rather than exponential. Force balance qualitatively changes the distribution and cannot be neglected \cite{vaneerd07}. Although experiments consistently find exponential tails in the distribution of boundary forces \cite{exptboundary}, experimental \cite{trush05,exptbulk} and numerical \cite{numerics} bulk distributions bend downward on a semilogarithmic plot, suggesting faster than exponential decay. 

%%%%%%%%%%%%%%%%%%%%%%%%%%%%%%%%%%%%%%%%%%%%%%%%%%%%%%%%%%%%%%%%%%%%%%%%%%%%%%%%%%%%%%%%%%%
{\em Force network ensemble}.---
%%%%%%%%%%%%%%%%%%%%%%%%%%%%%%%%%%%%%%%%%%%%%%%%%%%%%%%%%%%%%%%%%%%%%%%%%%%%%%%%%%%%%%%%%%%
Snoeijer's ensemble is composed of all ``force networks'', i.e.~sets of noncohesive contact forces on a {\em fixed granular contact network}, for which all $N$ grains are in {\em static force and torque balance} (e.g.~Fig.~\ref{fig:realrecip}a). For packings with more than a critical number of contacts per grain $z_c$, there exist many balanced force networks. $z_c = 4$ (3) for frictionless (frictional) 2D packings of disks \cite{snoeijer04}. All force networks on a contact network with the same $\sbar$ and local force and torque balance can be sampled uniformly by a series of Monte Carlo moves, termed ``wheel moves'' \cite{tighe05,vaneerd07}. Fig.~\ref{fig:realrecip}b gives an example. In the ensemble each force network has an equal {\em a priori} probability (a flat measure), in the spirit of the Edwards ensemble \cite{edwards}. We illustrate our approach in the specific case of the periodic frictionless triangular lattice of circular grains before expanding to frictional packings and different contact networks. 
%The idea is that the various balanced force networks capture the differences that would result from microscopic variations in grain shapes and positions. 
%Previous work has shown that force statistics from the ensemble are consistent with averaging over many molecular dynamics simulations of hard, but not perfectly rigid, grains. Likewise, force networks on disordered and ordered contact networks have qualitatively similar statistics.

As by construction the global stress tensor $\sbar$ is fixed \cite{stresstensor}, the extensive pressure in the system $\bigP$ is conserved,
\begin{equation} \label{eqn:fixedpressure}
\bigP = \sum_{i=1}^N p_i = \mathrm{const}\,.
\end{equation}
The sum runs over all grains and $p_i =  \mathrm{Tr}\,\hat{\sigma}_i/2$ is the pressure on grain $i$. We restrict ourselves to isotropic states, $\sbar \sim \mathbbm{1}$, so that $\bigP$ fully characterizes $\sbar$.
%All force networks on a particular contact network with the same $\sbar$ can be sampled uniformly by a series of Monte Carlo moves, termed ``wheel moves'' \cite{tighe05, vaneerd07}; Fig.~\ref{fig:realrecip}c gives an example. 
%There exists a basis of wheel moves for any contact network with $z>z_c$.
%There a move changes $12$ contact forces by $\delta f$. The six forces on a grain $i$ are incremented, while the six forces between nearest neighbors of $i$ are decremented. The strength $\delta f$ of the move is chosen from the interval $\delta f \in [-f_\mathrm{min}^{(i)}, f_\mathrm{min}^{(nn)}]$ with flat measure, where $f_\mathrm{min}^{(i)}$ is the smallest of the six forces on $i$ and $f_\mathrm{min}^{(nn)}$ is the smallest of the six forces between nearest neighbors of $i$. There is a wheel centered on each grain.

Every force network, regardless of its (dis)ordering, coordination number $z$, or friction coefficient $\mu$, has a corresponding reciprocal tiling. The systems in Fig.~\ref{fig:realrecip}a and 1c are a real and reciprocal pair; each grain corresponds to a tile. Each face of the tile corresponds to one of the grain's contact forces. The face is oriented at a $\pi/2$ rotation to the force $\f$, and its length is proportional to $|\f|$. Because the grain is in static force balance, the faces form a loop enclosing the tile \cite{blumenfeld02}. By Newton's third law, the tiles fit together with no gaps. 

Specifying the boundary forces on a packing establishes the boundaries of its corresponding tiling, and hence the tiling's total area. Fixing $\sbar$ in a periodic system has the same effect. Rearrangements of bulk forces, i.e.~the wheel moves, correspond to local exchanges of area among tiles. The {\em total} tiling area is unaltered by wheel moves, and therefore the area $\A$ is conserved. That is,
\begin{equation} \label{eqn:fixedarea}
\A = \sum_{i=1}^N a_i = \mathrm{const}\, .
\end{equation}
The sum runs over all tiles and $a_i$ is the area of tile $i$ \cite{area}. The conservation of $\A$ can be seen explicitly for the frictionless triangular lattice in Fig.~\ref{fig:realrecip}d. Area conservation is a {\em global} constraint that results from imposing {\em local} force balance. It holds for arbitrary force balanced packings in 2D with fixed $\sbar$ or boundary forces. It plays a crucial role in determining the statistics of local stresses. 

We scale the grain diameter in the triangular lattice such that the pressure on a grain is the sum of its $z=6$ normal forces. In these units the perimeter of a tile is equal to the pressure on the corresponding grain, making the pressure $p$ a convenient measure of local stress. Though the force distribution $\rho(f)$ is more commonly studied, we expect that $\rho(p)$ and $\rho(f)$ have similar tails.

%%%%%%%%%%%%%%%%%%%%%%%%%%%%%%%%%%%%%%%%%%%%%%%%%%%%%%%%%%%%%%%%%%%%%%%%%%%%%%%%%%%%%%%%%%%
{\em Entropy maximization}.---
%%%%%%%%%%%%%%%%%%%%%%%%%%%%%%%%%%%%%%%%%%%%%%%%%%%%%%%%%%%%%%%%%%%%%%%%%%%%%%%%%%%%%%%%%%%
\begin{figure}[tbp] 
\centering
\includegraphics[clip,width=0.95\linewidth]{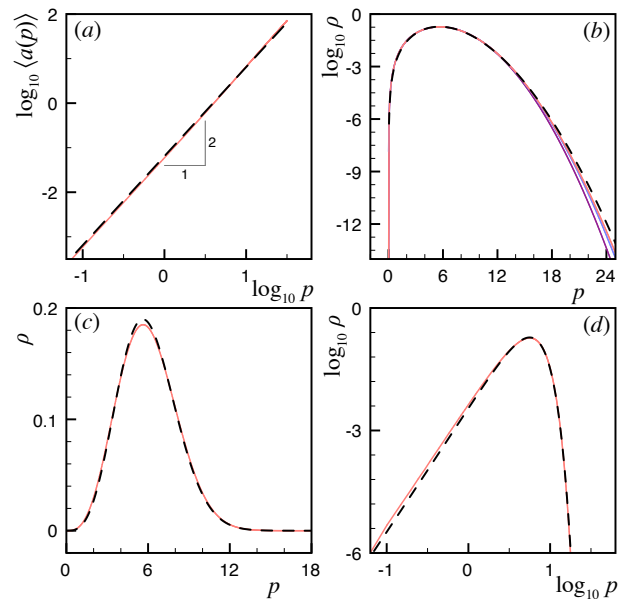}
\caption{(a) Fitted (dashed) and numerical (solid) average area of a tile with perimeter $p$.(b-d) Theoretical (dashed) and numerical (solid) pressure probability distributions for the frictionless triangular lattice with $N=1840$. The additional numerical curves in (b) are from $N=460$ and 115.  }
\label{fig:trilatt}
\end{figure}
Armed with the insight that the force network ensemble involves two conserved quantities, $\bigP$ and $\A$, we explore their implications for the statistics of local stresses. While previous work has incorporated the conservation of $\bigP$ or its equivalent, the conservation of $\A$ has heretofore been overlooked. We will show that the conservation of $\A$ has a dramatic effect on the force network statistics. 

We calculate the probability density $\rho(p)$ by maximizing entropy while conserving $\bigP$ and $\A$. Each force network corresponds to a set of pressures $\lbrace p_i \rbrace$, $i=1\ldots N$. We define $P(p)\omega(p)dp$ as the probability of finding a pressure $p$ in the interval $[p,p+\omega(p)dp)$, where $\omega(p)$ is the density of states for pressures. The entropy $S$ is the logarithm of the number of ways of constructing force networks with pressures $\lbrace p_i \rbrace$ consistent with $P(p)$. In the thermodynamic limit \cite{goddard04}
\begin{equation} \label{eqn:entropy}
S = -\int_0^\infty \bigl( P(p) \ln P(p) \bigr) \,\omega(p)dp\,.
\end{equation}
The experimentally accessible probability density $\rho(p)$ is related to $P(p)$ by $\rho(p)dp = P(p)\omega(p)dp$. 

It is important to note that weighting all force networks equally does not correspond to a flat measure on the pressures, i.e.~$\omega(p) \neq \mathrm{const}$. The contact forces $\lbrace \f_i \rbrace$ on a grain, $i=1\ldots z$, can be taken as coordinates of its state space. We demand that a grain explore only the regions of the space corresponding to force and torque balanced, noncohesive forces. We assume that, subject to these constraints and prior to imposing entropy maximization, the grain is equally likely to be in any of its allowed states; this amounts to neglecting correlations with neighboring grains \cite{spatial}. The result is a density of states that goes as $\omega(p) \propto p^\nu$. The value of $\nu$ depends on the grain's coordination number and the friction coefficient. For the frictionless triangular lattice, $\nu = z-3$.

The entropy is maximized subject to Eqs.~(\ref{eqn:fixedpressure}) and (\ref{eqn:fixedarea}), as well as normalization of $\rho(p)$.  This leads to
\begin{eqnarray} \label{eqn:constraints}
1 = \int_0^\infty \rho(p)\, dp \,,  &&
\pbar = \bigP/N = \int_0^\infty p\, \rho(p)\, dp\,,  \nonumber \\
\mathrm{and}\,\,\,\Abar = &\A/N& = \int_0^\infty \Abarp\, \rho(p) \, dp \,.
\end{eqnarray}
$\Abarp = \int  a\, \rho(a|p)\, da$ is the average area of a tile with perimeter $p$; $\rho(a|p)$ is the conditional probability a tile has area $a$ given perimeter $p$.
By the method of Lagrange multipliers the entropy-maximizing density subject to Eqs.~(\ref{eqn:constraints}) is 
\begin{equation} \label{eqn:genrho}
\rho(p) = Z^{-1} p^\nu \exp{(-\alpha p - \gamma \Abarp)}\, .
\end{equation}
Without the constraint on tiling area we would have $\gamma=0$ and an exponential tail: {\em Incorporating local force balance by means of the area constraint has qualitatively changed the form of the distribution}. The Lagrange multipliers $Z$, $\alpha$, and $\gamma$ are determined by substituting Eq.~(\ref{eqn:genrho}) in Eqs.~(\ref{eqn:constraints}). For frictionless systems a scaling argument shows that $\Abarp$ is quadratic in the thermodynamic limit \cite{scaling}; 
%footnote: Any single-grain state can be transformed in to a new one by multiplying its forces by $\lambda>0$. This scales the perimeter (area) of the corresponding tile by $\lambda$ ($\lambda^2$). Therefore $\rho(a|p) = \lambda^2 \rho(\lambda^2 a|\lambda p)$. Since for frictionless systems $a \geq 0$, it follows that $\Abarp \propto p^2$. 
we write $\Abarp = c \Abar (p/\pbar)^2$ and determine the constant $c$ from numerics. Thus the probability density $\rho(p)$ has a generically Gaussian tail, as was shown numerically for $\rho(f)$ \cite{vaneerd07}. 

We employ umbrella sampling \cite{vaneerd07} on a periodic frictionless triangular lattice with $N=1840$ to numerically determine $\rho(p)$. From the sampled $\Abarp$, shown in Fig.~\ref{fig:trilatt}a, we extract $c \approx 0.89$. Fig.~\ref{fig:trilatt}b-d contains the corresponding probability density of Eq.~(\ref{eqn:genrho}) and its numerical counterpart. Theory and numerics are in excellent agreement, even for $\rho(p)$ as low as $10^{-8}$. The slight discrepancies can be attributed to finite size effects and spatial pressure correlations: due to force balance, neighbors of large $p$ grains are more likely to be at high $p$ themselves. Thus large pressures are less entropically favorable than suggested by neglecting correlations.

%%%%%%%%%%%%%%%%%%%%%%%%%%%%%%%%%%%%%%%%%%%%%%%%%%%%%%%%%%%%%%%%%%%%%%%%%%%%%%%%%%%%%%%%%%%
{\em Frictional lattices}.---
%%%%%%%%%%%%%%%%%%%%%%%%%%%%%%%%%%%%%%%%%%%%%%%%%%%%%%%%%%%%%%%%%%%%%%%%%%%%%%%%%%%%%%%%%%%
\begin{figure}[tbp] 
\centering
\includegraphics[clip,width=0.95\linewidth]{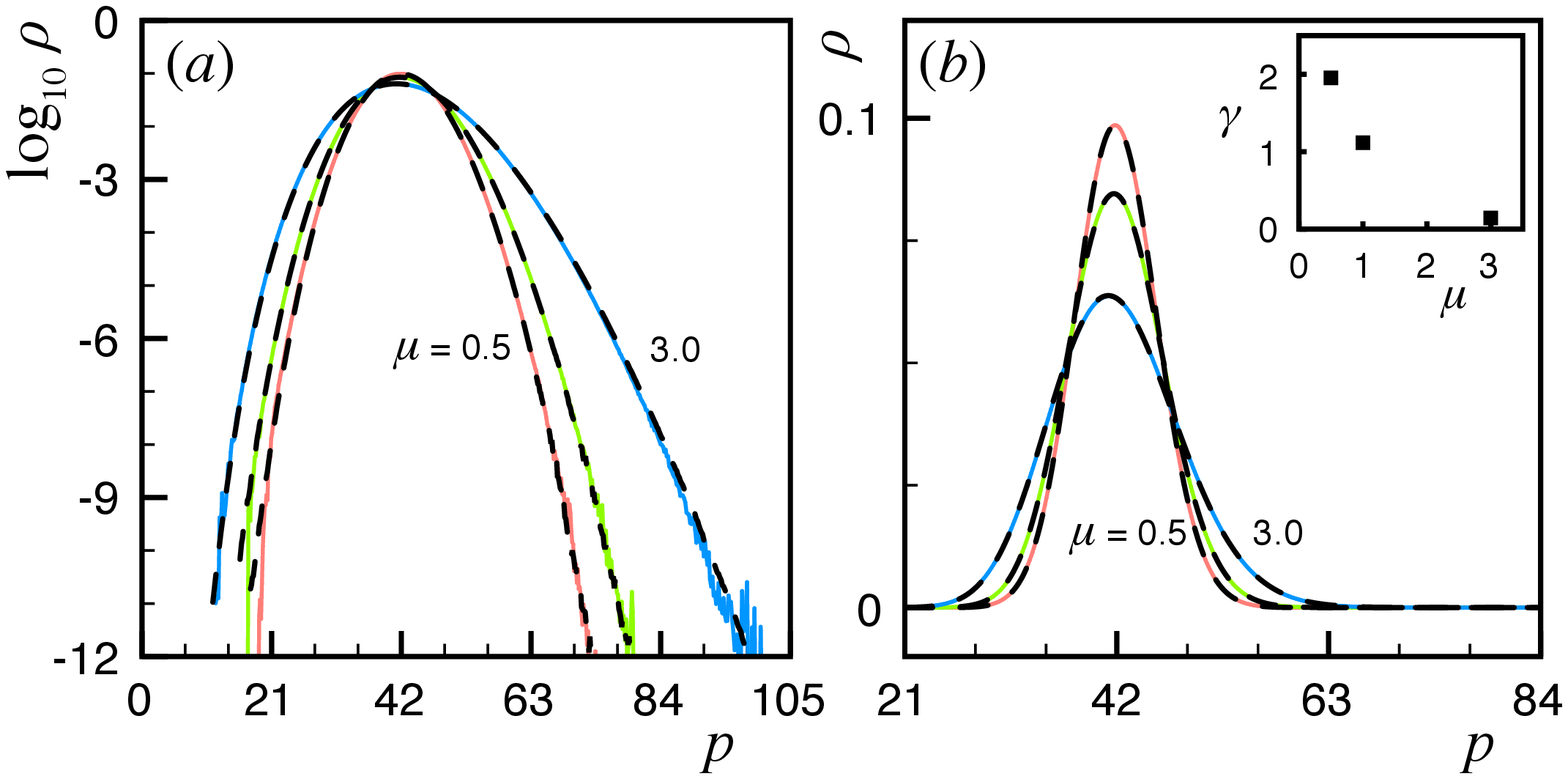}\\
\includegraphics[clip,width=0.95\linewidth]{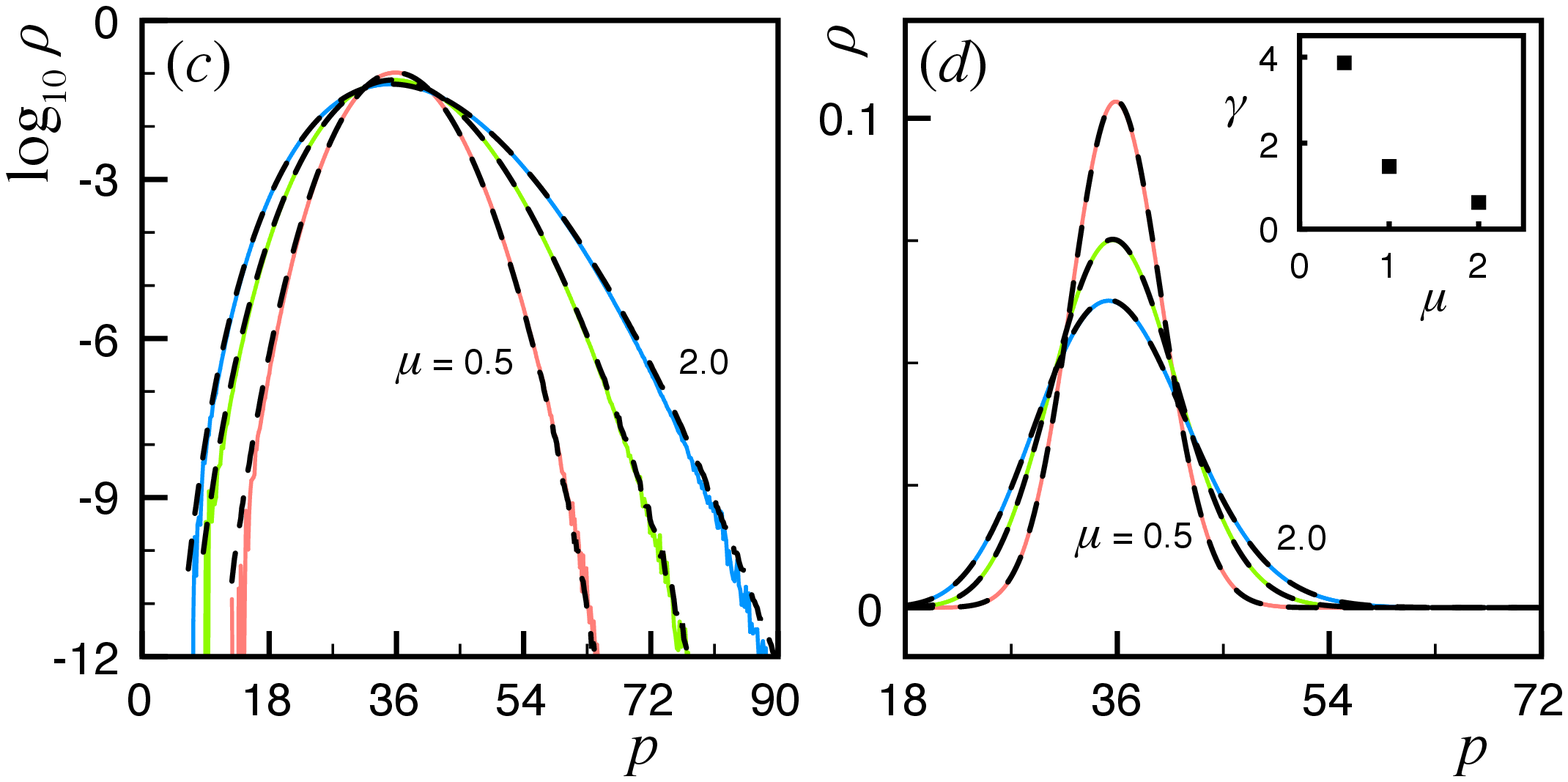}
\caption{(a,b) Theoretical (dashed) and numerical (solid) pressure probability distributions for 7-grain clusters in the frictional triangular lattice with $\mu = 0.1$, 0.5, 1.0, and 3.0. (c,d) Pressure distributions for 9-grain clusters in the frictional square lattice with $\mu = 0.5$, 1.0, and 2.0.  (b, d inset) $\gamma$ of Eq.~(\ref{eqn:genrho}) for for various friction coefficients $\mu$.}
\label{fig:friclatt}
\end{figure}
We now consider frictional triangular ($z=6$) and square ($z=4$) lattices. A system with friction coefficient $\mu$  permits  contact forces with a tangential component $t\leq \mu n$, $n$ being the normal component. This has two important consequences. 
% The first is to change the exponent $\nu$ in the density of states $\omega(p)$. The single-grain state space now has $2z$ coordinates $\lbrace n_i, t_i \rbrace$, $i=1\ldots z$, and torque balance is no longer satisfied trivially. By the same reasoning as above, $\nu=2z-4$. 
The first is that $\Abarp$ is not strictly quadratic. Friction permits tiles with area $a<0$, which occurs when tile faces overlap. Nevertheless, on dimensional grounds we expect $\Abarp \sim p^2$ for large $p$. In numerics, deviations from a quadratic form increase with $\mu$, but for all frictional systems we have studied quadratic scaling holds for $p\gtrsim \pbar$. Hence Eq.~(\ref{eqn:genrho}) still yields Gaussian tails. Secondly, we find that friction increases spatial correlations \cite{spatial}. 
%footnote: We have verified that the spatial structure factor $S(q)=\langle |p_\underline{q}|^2 \rangle$ is nearly flat for the frictionless case, indicating weak spatial correlations. As $\mu$ is increased, $S(q)$ develops an increasingly negative slope, corresponding to spatially smoother and more correlated stresses.
Consequently, as in Ref.~\cite{henkes07}, we coarse-grain and study clusters of $m=7$ (9) grains and $k=30$ (24) contacts on the triangular (square) lattice. The frictional clusters have exponent $\nu=2k - 3m -1$ in their density of states. We find  $\Abarp$ for a cluster deviates much less from quadratic behavior than its single-grain counterpart.

Lacking the exact form of $\Abarp$ for frictional systems, we determine the Lagrange multipliers satisfying Eqs.~(\ref{eqn:constraints}) using the numerically sampled $\Abarp$. Theory and numerics are again in excellent agreement, as seen in Fig.~\ref{fig:friclatt}. 

%%%%%%%%%%%%%%%%%%%%%%%%%%%%%%%%%%%%%%%%%%%%%%%%%%%%%%%%%%%%%%%%%%%%%%%%%%%%%%%%%%%%%%%%%%%
{\em Infinite friction}.---
%%%%%%%%%%%%%%%%%%%%%%%%%%%%%%%%%%%%%%%%%%%%%%%%%%%%%%%%%%%%%%%%%%%%%%%%%%%%%%%%%%%%%%%%%%%
As the Lagrange multiplier $\gamma$ tends towards zero with increasing $\mu$ (Fig.~\ref{fig:friclatt}b and d, insets), we investigate the limit $\mu \rightarrow \infty$. For finite friction and circular grains, normal and tangential forces are coupled through the force balance constraints on each grain and the Coulomb constraint on each force. In the infinite friction limit the Coulomb constraint has no effect. For the triangular lattice there are three distinct contacts per grain, and it is possible to choose a set of tangential forces $\lbrace t_i \rbrace$ to balance force and torque on each grain regardless of the normal forces $\lbrace n_i \rbrace$. The only constraints on the $\lbrace n_i \rbrace$ are positivity, $n_i>0$, and fixed total pressure $\mathcal{P}$. This leads directly to a Boltzmann distribution $\rho(n) = \langle n \rangle^{-1}\exp{(-n/\langle n \rangle)}$ with $\langle n \rangle = \mathcal{P}/zN$. In contrast, for systems with $z-z_c < 3$, such as the square lattice, the normal and tangential forces remain strongly coupled through force balance even for infinite friction. We have confirmed numerically that in the infinite friction limit the Boltzmann distribution holds for the triangular lattice, and that the normal force and pressure distributions in the square lattice remain Gaussian.

%%%%%%%%%%%%%%%%%%%%%%%%%%%%%%%%%%%%%%%%%%%%%%%%%%%%%%%%%%%%%%%%%%%%%%%%%%%%%%%%%%%%%%%%%%%
{\em Boundary forces}.---
%%%%%%%%%%%%%%%%%%%%%%%%%%%%%%%%%%%%%%%%%%%%%%%%%%%%%%%%%%%%%%%%%%%%%%%%%%%%%%%%%%%%%%%%%%%
To this point we have imposed a flat measure on ensembles of periodic force networks. We have also numerically investigated the frictionless triangular lattice with a boundary and subject to a flat measure on the {\em boundary} forces \cite{blumenfeld07}. This produces by prescription a Boltzmann distribution of boundary forces, reminiscent of experiment \cite{exptboundary}. Nevertheless, we find that the force and pressure distributions in the bulk, defined as grains at least six layers from the boundary, have Gaussian tails. This simple example demonstrates that a boundary distribution may not provide direct information about the bulk distribution.

%%%%%%%%%%%%%%%%%%%%%%%%%%%%%%%%%%%%%%%%%%%%%%%%%%%%%%%%%%%%%%%%%%%%%%%%%%%%%%%%%%%%%%%%%%%
{\em Conclusion}.---
%%%%%%%%%%%%%%%%%%%%%%%%%%%%%%%%%%%%%%%%%%%%%%%%%%%%%%%%%%%%%%%%%%%%%%%%%%%%%%%%%%%%%%%%%%%
We have derived an analytic expression for the distribution of pressures in the force network ensemble in 2D and found excellent agreement with numerics. Distinct from previous studies, we incorporate {\em two} conserved quantities, a total pressure and the area of a reciprocal tiling. The latter is a direct consequence of force balance on the grain scale, and we conclude that this is crucial to understand the statistics of local forces in granular media. As a result, large stresses obey Gaussian statistics. This observation is robust to changes in the contact network, the finite friction coefficient, and the imposed measure.

We have {\em not} addressed the distribution at the unjamming transition, which could have a signature in the local stress statistics. Marginally rigid packings cannot be studied within the force network ensemble. Similarly, our results are restricted to two dimensions. A na\"ive extension of the reciprocal tiling to 3D suggests $\rho(p) \sim e^{-p^\delta}$ with $\delta=3/2$, while numerics find $\delta \approx 1.7$ \cite{vaneerd07} within the force network ensemble. The discrepancy may be the result of stronger spatial correlations than in 2D, where coarse-graining suffices, or it may signal new physics. 

Importantly, along with recent experiments \cite{trush05,exptbulk}, our results give serious cause to doubt that exponential statistics are a generic property of jammed granular matter. At the very least, more work is needed to distinguish bulk and boundary phenomena and to clarify why measured boundary forces show exponential statistics.

% We have derived an expression for the distribution of pressures in the force network ensemble and confirmed it with numerics. Distinct from previous work, we incorporate {\em two} conserved quantities, a total pressure and the area of a reciprocal tiling. The latter is a direct consequence of force balance on the grain scale. Thus force balance is crucial to understanding the statistics of local forces in granular media. Due to these conserved quantities, large stresses obey Gaussian statistics. This observation is robust to changes in the contact network, the finite friction coefficient, and the imposed measure. 
% Though our approach is not exact, the influence of neglected spatial correlations can be diminished by coarse-graining, while the pressure and area constraints are unaffected. Obvious questions regard packings at marginal rigidity, which we cannot access within the force network ensemble, as well as 3D packings. A na\"ive extension of the reciprocal tiling to 3D suggests $\rho(p) \sim e^{-p^\delta}$ with $\delta=3/2$, while numerics find $\delta \approx 1.7$ \cite{vaneerd07}. More generally, along with recent experiments \cite{trush05,exptbulk}, our results give reason to doubt that exponential statistics are a generic property of jammed granular matter. At the very least, more work is needed to distinguish bulk and boundary phenomena. 

\enlargethispage{7mm}
%%%%%%%%%%%%%%%%%%%%%%%%%%%%%%%%%%%%%%%%%%%%%%%%%%%%%%%%%%%%%%%%%%%%%%%%%%%%%%%%%%%%%%%%%%%
{\em Acknowledgments}.---
%%%%%%%%%%%%%%%%%%%%%%%%%%%%%%%%%%%%%%%%%%%%%%%%%%%%%%%%%%%%%%%%%%%%%%%%%%%%%%%%%%%%%%%%%%%
We thank Wouter Ellenbroek, Zorana Zeravcic, Martin van Hecke, and Wim van Saarloos for helpful conversations. BPT acknowledges support from the physics foundation FOM and the hospitality of the Aspen Center for Physics, where part of this work was done. 

%\bibliographystyle{apsrev}
%\bibliography{thesisbib}

\begin{thebibliography}{21}
\bibitem{stresstensor} $\sigma_{\alpha \beta} = \sum_{i,j} f^{(i,j)}_\alpha r^{(i,j)}_\beta/2V$. The sum runs over all contacting grain pairs $(i,j)$ in volume $V$. $\f^{(i,j)}$ is the force from $j$ on $i$ and $\bond^{(i,j)}$ is the vector between their centers.
\bibitem{spatial}
The structure factor $S(q)=\langle |p_{\boldsymbol{q}}|^2 \rangle$ is nearly flat for the frictionless triangular lattice, indicating weak spatial correlations. As $\mu$ is increased, $S(q)$ develops an increasingly negative slope, corresponding to more correlated stresses.
\bibitem{scaling} Any single-grain state can be transformed into a new one by multiplying its forces by $\lambda>0$. This scales the perimeter (area) of the corresponding tile by $\lambda$ ($\lambda^2$). Therefore $\rho(a|p) = \lambda^2 \rho(\lambda^2 a|\lambda p)$. Since for frictionless systems $a \geq 0$, it follows that $\Abarp \propto p^2$.
\end{thebibliography}

%\bibitem{area} $a_i = \frac{1}{2}\hat{z}\cdot\sum_{j=1}^{z_i-2}\sum_{k=j+1}^{z_i-1} (\f^{(i,j)} \times \f^{(i,k)})$. The indices $j$ and $k$ label the $z_i$ neighbors of grain $i$, proceeding around the grain in a right-hand fashion. 
%
%\bibitem{scaling} Any single-grain state can be transformed in to a new one by multiplying its forces by $\lambda>0$. This scales the perimeter (area) of the corresponding tile by $\lambda$ ($\lambda^2$). Therefore $\rho(a|p) = \lambda^2 \rho(\lambda^2 a|\lambda p)$. Since for frictionless systems $a \geq 0$, it follows that $\Abarp \propto p^2$. 
%
%\bibitem{spatial}
%We have verified that the structure factor $S(q)=\langle |p_\underline{q}|^2 \rangle$ is nearly flat for the frictionless case, indicating weak spatial correlations. As $\mu$ is increased, $S(q)$ develops an increasingly negative slope, corresponding to spatially smoother and more correlated stresses.

\end{document}